\documentclass[a4paper,fleqn,usenatbib]{mnras}

\usepackage[T1]{fontenc}
\usepackage{ae,aecompl}

\usepackage{graphicx}	
\usepackage{amsmath}	
\usepackage{amssymb}	
\usepackage{subfig}
\usepackage{arydshln}

\title[SDSS IV MaNGA: Full spectroscopic bulge-disc decomposition]{SDSS IV MaNGA: Full spectroscopic bulge-disc decomposition of MaNGA early-type galaxies}
\author[M. Tabor et al.]{Martha Tabor,$^{1}$\thanks{E-mail: martha.tabor@nottingham.ac.uk}
Michael Merrifield,$^{1}$
Alfonso Arag\'{o}n-Salamanca,$^{1}$
\newauthor
Amelia Fraser-McKelvie,$^{1}$
Thomas Peterken,$^{1}$
Rebecca Smethurst,$^{1,2}$
\newauthor
Niv Drory,$^{3}$
and Richard R. Lane,$^{4,5}$
\\
$^{1}$School of Physics and Astronomy, University of Nottingham, University Park, Nottingham, NG7 2RD, UK\\
$^{2}$Department of Astrophysics, University of Oxford, Denys Wilkinson Building, Keble Road, Oxford OX1 3RH, UK\\
$^{3}$McDonald Observatory, The University of Texas at Austin, 1 University Station, Austin, TX 78712, USA\\
$^{4}$ Instituto de Astrof\'{i}sica, Pontificia Universidad Cat\'{o}lica de Chile, Av. Vicu\~na Mackenna 4860, 782-0436 Macul, Santiago, Chile \\
$^{5}$Millennium Institute of Astrophysics, Av. Vicu\~na Mackenna 4860, 782-0436 Macul, Santiago, Chile
}

\date{Accepted XXX. Received YYY; in original form ZZZ}

\pubyear{2015}

\begin{document}
\label{firstpage}
\pagerange{\pageref{firstpage}--\pageref{lastpage}}
\maketitle
\begin{abstract}

By applying spectroscopic decomposition methods to a sample of MaNGA early-type galaxies, we separate out spatially and kinematically distinct stellar populations, allowing us to explore the similarities and differences between galaxy bulges and discs, and how they affect the global properties of the galaxy. We find that the components have interesting variations in their stellar populations, and display different kinematics. Bulges tend to be consistently more metal rich than their disc counterparts, and while the ages of both components are comparable, there is an interesting tail of younger, more metal poor discs. Bulges and discs follow their own distinct kinematic relationships, both on the plane of the stellar spin parameter, $\lambda_{R}$, and ellipticity, $\epsilon$, and in the relation between stellar mass, $M_\ast$, and specific angular momentum, $j_\ast$, with the location of the galaxy as a whole on these planes being determined by how much bulge and disc it contains. As a check of the physical significance of the kinematic decompositions, we also dynamically model the individual galaxy components within the global potential of the galaxy. The resulting components exhibit kinematic parameters consistent with those from the spectroscopic decomposition, and though the dynamical modelling suffers from some degeneracies, the bulges and discs display systematically different intrinsic dynamical properties. This work demonstrates the value in considering the individual components of galaxies rather than treating them as a single entity, which neglects information that may be crucial in understanding where, when and how galaxies evolve into the systems we see today.

\end{abstract}




\section{Introduction}

A crucial step in understanding the formation history of a galaxy is knowing where and when its constituent stellar
populations were formed. As a first approximation of this, photometric bulge-disc decompositions have long been 
used as a way of separating out the central 
and extended components in a galaxy. 
However, detailed understanding of these components, both in terms of the stellar populations that they are built from and their kinematic properties, is still lacking. 

With the development of integral field spectroscopy (IFS) surveys such as SAURON
\citep{deZeeuw2003}, ATLAS$^{\rm 3D}$ \citep{Cappellari2011}, SAMI
\citep{Bryant2015}, CALIFA \citep{Sanchez2012} and the MaNGA survey \citep{Bundy2015}, these questions are beginning to be addressed. \citet{Tabor2017} demonstrated a new method of spectroscopic decomposition, showing that photometric bulges and discs of early-type galaxies (ETGs) represent individual stellar populations with distinct kinematics. Through various other methods, from simply masking out contaminated regions in order to look at spectral indices in bulge and disc regions \citep{Fraser2018}, to more detailed processes, such as \citet{Johnston2014}, \citet{Taranu2017}. \citet{Coccato2018} and \citet{Rizzo2018}, other studies have also demonstrated the existence of two distinct components containing different stellar populations, the relative ages and metallicities of which appear to vary intriguingly with mass and, potentially, environment.

Work by the ATLAS$^{\rm 3D}$ team demonstrated the effectiveness of using IFS data to extract and parameterise the kinematics of galaxies. The introduction of the angular momentum parameter, $\lambda_{R}$, revealed the large number of early-type galaxies with disc-like kinematics, and led to the new classification of ETGs into fast and slow rotators \citep{Emsellem2007}, or regular and non-regular rotators, where the latter group includes slow rotators and kinematically decoupled cores \citep{Cappellari2007}. The prevalence of these discs makes the question of what they are and how they form all the more interesting.

Dynamical modelling of the stellar kinematics can provide a useful way of quantifying this kinematic information. By using an anisotropic generalisation of the semi-isotropic axisymmetric Jeans formalism \citep{Cappellari2008}, \citet{Cappellari2016} demonstrated how, using anisotropy parameters extracted though modelling, the differences and similarities across the early-type galaxy population can be explored. However, the dynamical models used so far to describe the galaxy kinematics treat the galaxy as a single dynamical component. With the increasing evidence for the presence of two kinematically distinct populations within many early-type galaxies, allowing for a second dynamical component is not only a more accurate representation of a galaxy, but can also provide an additional way to explore the dynamics of the individual components. 
 
The specific angular momentum, $j_\ast$, has been shown to correlate well with stellar mass, $M_\ast$. Looking at where galaxies lie on the $M_\ast - j_\ast$ relation is another useful way of understanding their evolution. The scatter in the relation has been shown to correlate with morphological properties such as S\'{e}rsic index \citep{Sersic1963} and bulge-to-total ratio \citep{Fall1983,Cortese2016,Romanowsky2012}, indicating that where galaxies sit on the relation is determined by the relative contribution of their bulge and disc, each of which could then be placed on their own separate $M_\ast - j_\ast$ relation. Being able to extract the kinematics of the components and determine where they lie on this plane is therefore important in understanding the role of each component in the evolution of the galaxy as a whole.

In this work, we aim to explore the role bulges and discs play in determining the global properties of galaxies, which have often been historically treated as a single entity. By separating stellar components that are spatially and kinematically segregated, with distinct chemical compositions and age distributions, we can extract observational kinematic properties such as $\lambda_{R}$ and $j_\ast$, dynamical properties through modelling, and stellar population parameters, such as age and metallicity, for bulges and discs separately. Through this decomposition, we can begin to understand the effects each component has on the galaxy as a whole, and understand their separate formation and evolution history. 

To this end, we have spectroscopically decomposed a sample of MaNGA early-type galaxies into bulges and discs. The method presented in \citet{Tabor2017} was demonstrated on a small sample of galaxies, allowing the fits to be heavily monitored. Applying this technique to a larger sample of galaxies allows us to explore the feasibility of applying the decomposition method to large IFS surveys in an automated manner. 

Without discriminating by kinematics, we apply the decomposition to all galaxies with a clear bulge and disc found during a photometric decomposition, extracting and analysing kinematics of the bulge and disc separately. Both in order to check the validity of the spectroscopic decomposition, and to determine how these components correspond to dynamical systems sitting within the global potential of the galaxy, we use dynamical modelling methods to determine dynamical properties of the individual components. 
The decomposition method also allows the extraction of the stellar populations of the components which we present and compare to results found using other methods of age and metallicity determination. 

With mergers, dissipative collapse and secular evolution all leaving their fingerprints in the orbits, chemical composition and age distribution of the stars, this combining of the ages and metallicities with the kinematics of the individual components could be the key to determining where and when these processes dominate. 

The paper is structured as follows: in section \ref{dat} we present the data and sample selection, in section \ref{method} we outline the method used for the spectroscopic decomposition, with the resulting stellar populations presented in section \ref{pops} and the kinematics in section \ref{kins}, along with the dynamical modelling. Finally in section \ref{diss} we discuss the implications of the results obtained, with key conclusions presented in section \ref{conc}.

\section{Data and Galaxy Sample} \label{dat}
 
 \subsection{The MaNGA Survey}
In this work we use data from the Mapping Nearby  Galaxies  at  Apache  Point  Observatory (MaNGA) survey \citep{Bundy2015,Drory2015}, which aims to obtain 
integral field spectroscopy for 10,000 nearby galaxies in the redshift range $0.01 < z < 0.15$ by 2020 \citep{Yan2016b}. It is a project of SDSS-IV \citep{Blanton2017}, using the 2.5m telescope at Apache Point Observatory \citep{Gunn2006} and BOSS spectrographs \citep{Smee2013}. The data cubes cover a wavelength range of 3600--10300\AA, with
a spectral resolution R $\sim2000$ and a spatial resolution of 2.5". The survey aims to observe a luminosity-dependent, volume-limited sample of galaxies above a stellar mass of $10^{9}$, with a roughly flat $\log( M_\ast$)  distribution \citep{Law2015,Wake2017}
There are various integral field unit (IFU) sizes with a distribution optimised to match the size and density of the targets, ranging from 127-fibre 
IFUs of 32 arcseconds to 19-fibre IFUs of 12 arcsecond diameter. The raw data are reduced by a data reduction pipeline \citep{Law2016,Yan2016a} and made available as a final data-cube, comprising a spectrum at each spatial location, for each galaxy.

\subsection{Sample Selection}

To select a sample of early-type galaxies, we use the Galaxy Zoo 2 catalogue \citep{Lintott2011, Willett2013}, which provides statistics on galaxy characteristics voted for by the public for SDSS galaxies.
By selecting all objects with `smooth' vote fraction $>$ 0.7 \citep{Willett2013}, with the vote fractions debiased to account for redshift-dependent bias as described in \citet{Hart2016}, we rule out spiral and irregular galaxies while keeping 
all early-type galaxies including lenticulars and ellipticals. We also remove galaxies with bars (bar fraction $>$ 0.7) so as to avoid more complicated kinematics, leaving 821 galaxies in our sample.

The decomposition requires high signal-to-noise spectra. We therefore bin the data-cubes spatially as described in section \ref{decomp}. In order to get an idea of the spatial variations across the galaxy we decided to only include galaxies with more than 30 bins. To ensure that the properties of both components can be effectively extracted we also remove galaxies where either component has fewer than 5 bins in total that contribute more than 30\% of the total flux in that bin, where the components are defined according to the photometric decomposition described in section \ref{phot}. This results in galaxies with one 
very faint photometric component not being decomposed, effectively taking a cut in bulge-to-total ratio as all 
of these galaxies have either B/T less than 0.1 or greater than 0.9. While these criteria exclude many of the less luminous, lower mass galaxies, it leaves us with a core of 302 high signal-to-noise cubes with clear photometric bulge and disc components.

\section{The method} \label{method}

\subsection{Photometric decomposition} \label{phot}

To obtain the light distribution of the bulges and discs of each galaxy, we performed a photometric bulge--disc decomposition on $r$-band SDSS NASA Sloan Atlas images, with sky subtraction performed as described in \citet{Blanton2011}. Using GALFIT \citep{Peng2002}, we fit an exponential disc and a S\'{e}rsic bulge, leaving S\'{e}rsic index, magnitudes, position angles and scale radii as free parameters, using NSA photometric parameters as initial estimates. To then obtain the bulge and disc fraction across the MaNGA data-cube, we repeat the decomposition on the MaNGA un-binned data-cube, averaged across the whole wavelength range to form a single image, fixing the S\'{e}rsic index, position angles, axis ratios and scale radii to those obtained in the previous decomposition. 
GALFIT produces an image of the bulge and disc components, convolved with the PSF of MaNGA data, which can then be binned into the same bins as the full data-cube (see section \ref{decomp}), giving the fraction of light in the bulge and disc within each spatial bin.

\subsection{Spectroscopic decomposition} \label{decomp}
\begin{figure*}
\begin{center}
	\includegraphics[width=\textwidth]{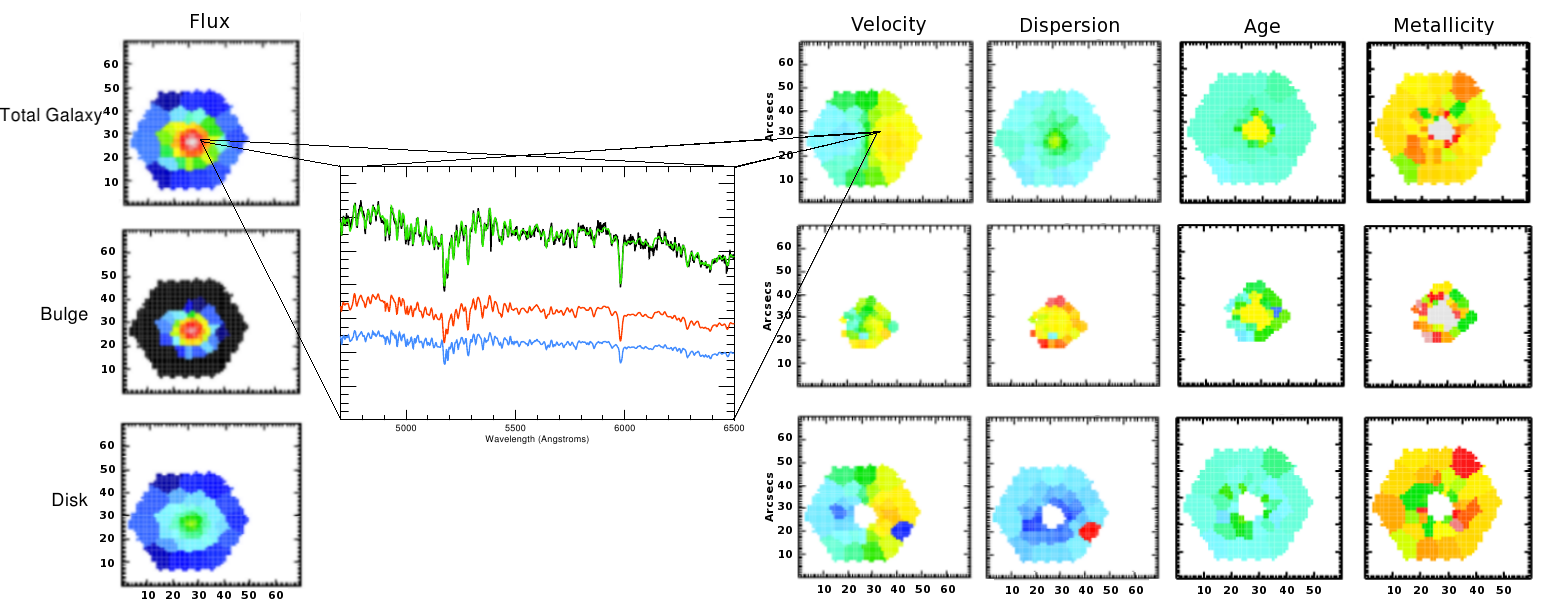} 
        \caption{Diagram of the decomposition process on a sample MaNGA data cube described in Section \ref{decomp}. On the left is the binned flux maps of the total galaxy, the bulge and the disc, and on the right the maps of velocity, dispersion, age and metallicity arising from the spectroscopic decomposition. The galaxy spectrum for a single bin in the galaxy is shown in black, the decomposed bulge and disc spectra in red and blue, and the sum of the bulge and disc in green.\label{fig:schem}}
 \end{center}
 \end{figure*}

In this paper we utilise the method presented in \citet{Tabor2017}. By fitting spectra at each spatial point in the galaxy with two separate 
model spectra, composed of a linear combination of single stellar population templates, and
constraining the models such that their relative surface brightness 
follows that provided by a previous photometric decomposition, the kinematics, ages and metallicities of each component 
can be extracted. 
Figure~\ref{fig:schem} gives a schematic of the decomposition process and the basic steps are described below.

The data-cube is first spatially binned to a signal-to-noise ratio (SNR) of 30 using the Voronoi binning method of \citet{Cappellari2003}. This high SNR was found to be necessary in order to accurately reproduce kinematics and stellar populations of simulated data, as described in \citet{Tabor2017}. To make the decomposition process more efficient, for especially luminous galaxies which still had large numbers of Voronoi bins after binning to SNR=30, we increase the SNR until the number of bins is less than 100. This allows us to retain spatial information without spending huge amounts of time decomposing a single galaxy. The kinematic decomposition is then performed by fitting each bin in the 
galaxy with two model spectra. For this we used the Python 
version of the penalised pixel-fitting 
code\footnote{Available from \url{http://purl.org/cappellari/software}} [pPXF; \citet{Cappellari2004}] which, from version 5.1, allows multiple components with different populations and kinematics to be fitted to spectra simultaneously \citep[see ][]{Johnston2013a}.

To appropriately constrain the fit and to ensure that the bulge and disc were assigned the correct kinematics, we use a model bulge and disc spectrum in the decomposition. These are built by averaging the bins where B/T > 0.7 for the bulge and B/T < 0.3 for the disc, where B/T is the bulge-to-total ratio in a given bin, and running pPXF on these two spectra. The resulting model spectra, built from SSP models, are then used as as templates for the future two component fits.
The flux ratio of the two components is also constrained to that obtained during the photometric decomposition as described in Section \ref{phot}. 

Finally, to ensure the true best fit to the galaxy spectrum is found each time, for each Voronoi bin we take the best fit to the spectrum across a $5\times5$ grid of bulge and disc velocities ranging from $-350\,$ km s$^{-1}$ to $350\,$ km s$^{-1}$. 
By constraining the velocities of each component to be within the pixel size of the grid, pPXF is forced to cover the whole velocity range, allowing the true chi-squared minimum to be selected, representing the best-fitting bulge and disc velocity.
This provides measurements of line-of-sight velocity and velocity dispersion for the bulge and disc at each 
spatial bin in the galaxy. 

Once the kinematics have been obtained we then repeat the decomposition, using the full library of SSP models but constraining the kinematics to those found in the previous decomposition. We used the single stellar 
population templates of \citet{Vazdekis2010}, providing high-resolution
spectra for stellar populations with metallicities ranging from
$-2.32$ to $0.22$ and ages from $0.06\,{\rm ~Gyr}$ to $17.8\,{\rm ~Gyr}$, resulting in 
a total catalogue of 155 spectra. 
We found that the stellar populations obtained using this iterative process were similar to those obtained when fitting the stellar populations and kinematics simultaneously, but the kinematics were improved due to the trade-off between the components being reduced.

To make sure the decompositions are truly separating out bulges and discs, we enforce some cuts on the data. 
Firstly, in order to avoid including noise, we include bins only in
regions where a component contributes at least 30\% of the total flux.  We also do not include bins where the velocity dispersion of the component is at the far limit of $\sigma=10$ or $\sigma=500kms^{-1}$ set during the pPXF fit, an indication that pPXF has failed to find the true best fit.
The kinematic parameters discussed in the rest of this paper are therefore determined for each component in the bins which contribute at least 30\% of the overall flux and are within 1 component effective radius. 

Secondly, partly because not all galaxies in the sample have two kinematic components and partly due to the complexities inherent
in performing the decomposition
on large numbers of galaxies, the decomposition occasionally fails to effectively fit the kinematics. To account for this, we remove any
galaxies where the decomposed disc component has a stellar spin parameter, $\lambda_{Re}$, which falls below the fast/slow rotator divide, where $\lambda_{R}$ is defined as
\begin{equation} \label{eq:lam}
\lambda_{R}=\frac{\sum {F_{i}  R_{i}  \mid V_{i} \mid}} {\sum F_{i} R_{i} \sqrt{V_{i}^2 + \sigma_{i}^2}},
\end{equation}
where $R_i$, $F_i$,$V_i$ and $\sigma_i$ correspond to the radius, flux, line-of-sight velocity and line-of-sight velocity dispersion of the $i^{th}$ Voronoi bin \citep{Emsellem2007}. Fast rotators are then classed as those with 
\[ 
\lambda_{Re} > 0.08 + \epsilon / 4 \quad \text{or} \quad \epsilon > 0.4,
\]
calculated within one effective radius, where $\epsilon$ is taken from the photometric decomposition described in Section \ref{phot}. Removing these galaxies after the decomposition 
rather than making an initial cut on $\lambda_{Re}$ insures we are not missing galaxies with faint rotating components not evident
in the global kinematic measurements. This process removes 30 galaxies, leaving a remaining sample of 272 galaxies.

\subsection{Aperture effects}
Determining parameters within 1$\rm R_e$ of the components allows a comparison of components in a similar way to studies done across the galaxy population. However, parameters such as $\lambda_{R}$ and $j_\ast$ have been shown to depend on the choice of radius within which they are calculated. This may be due to the bulge and disc components dominating different regions, leading to lower angular momentum in the bulge-dominated central regions of galaxies, but can also be the case even within a pure disc due to the shape of the velocity profile.
Comparing the angular momentum of the different components, which dominate in different regions of the galaxy, is therefore subject to biases depending on where the components are in the galaxy, and the radius over which they are calculated. This makes comparisons of the intrinsic properties of the components difficult. 

To allow a meaningful comparison we therefore also determine the parameters only using bins where we have information for both components, i.e. bins where both the bulge and disc contribute greater than 30\% of flux, and include these as additional figures in the following sections. Because we also do not include bins where the decomposition has failed, some galaxies in the sample will not have any bins in this overlapping region, which is why the sample size is reduced in these figures. The distribution of the fraction of bins which lie in these overlapping regions is large, ranging from 0 to 0.73, with a mean fraction of 0.26, and does not depend on mass or the effective radii of the components, therefore demonstrating that we do not introduce any biases when determining the parameters.

\section{Stellar Populations} \label{pops}

\begin{figure*}
    \centering
    \includegraphics[width=\textwidth]{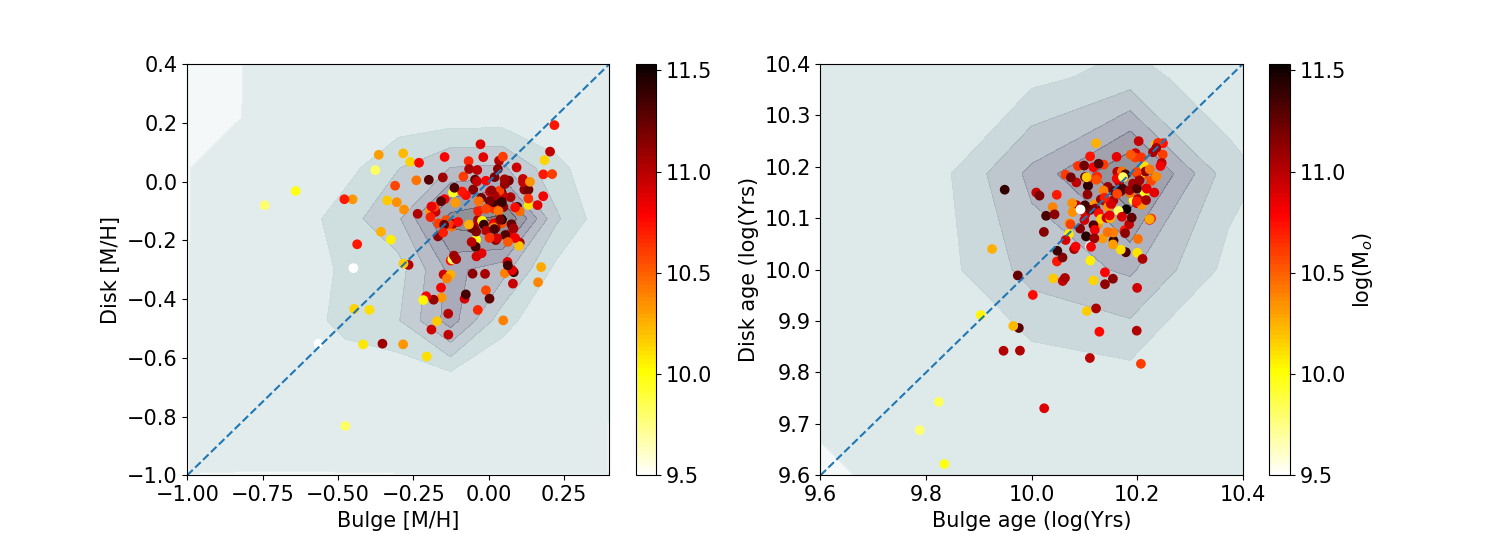}
    \caption{Age (right) and metallicity (left) for the bulge and disc components, coloured by mass. Contours (greyscale) show the density of points, weighted by their representation in the MaNGA sample. The dashed line is where bulge and disc parameters are equal.}
    \label{fig:SSP}
\end{figure*}

Figure \ref{fig:SSP} shows the flux-weighted ages and metallicities of the two components, coloured by galaxy stellar mass. Due to the flat mass selection of the MaNGA sample, higher mass galaxies are over-represented. The contours shown are therefore weighted by the representation of each galaxy in the sample, assigning higher mass galaxies less weight than lower mass galaxies. While the ages of the bulges and discs are comparable, there is a clear offset in metallicity towards bulges being more metal rich, and an interesting tail reaching to younger, more metal poor discs relative to their companion bulges. 
We see very similar distributions in the populations when only including overlapping bulge and disk regions, demonstrating that any differences in the components are not simply due to radial variations in the galaxies.

When compared to galaxies of a similar mass, these results are broadly consistent with results found by \citet{Fraser2018} who separate S0 galaxies into bulge and disc-dominated regions, and compare their ages and metallicities extracted from spectral indices. While our measurements of metallicity agree very well, we tend to find slightly older ages. This is expected when comparing spectral index measurements to results from full spectral fitting as done here, since the age-sensitive indices used by \citet{Fraser2018} lie at the blue end of the spectrum and are therefore more sensitive to younger populations, while full spectral fitting also takes into account features sensitive to the older stellar populations at the red end of the spectrum.

The similarity in the ages and metallicities of bulges and discs suggests a coupling between the components, which have evidently evolved together, a scenario previously suggested by \citet{Laurikainen2010}. The uniformly old stellar ages show that, as expected for such featureless systems, star formation ceased long ago, though the tail to younger, more metal poor discs indicates that on some occasions discs may have either been quenched later than their companion bulges, or possibly experienced some rejuvenation. We also looked for any trends with bulge-to-total ratio, but did not find any dependence.

\begin{figure}
    \centering
    \includegraphics[width=0.9\linewidth]{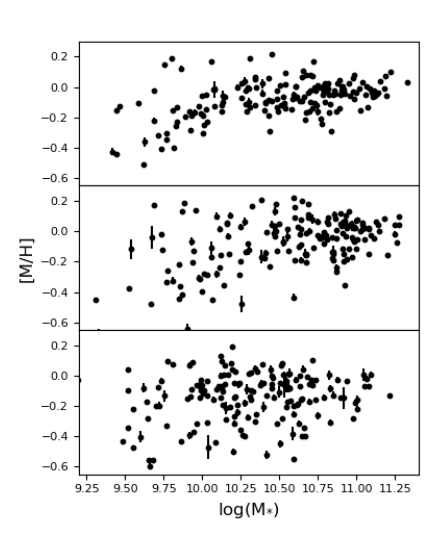}
    \caption{Mass--metallicity relation for the global galaxy (top) bulge (centre) and disc (bottom). Component masses are approximated by $M_{*} \times B/T$ and and $M_{*} \times (1-B/T)$ for the bulge and disc respectively. This assumes the same mass-to-light ratio in the components which is an approximation, but a reasonable one given their similar stellar populations. One sigma error-bars are shown but are often too small to be visible.}
    \label{fig:met}
\end{figure}

Figure \ref{fig:met} shows the mass-metallicity relation for the global galaxy and for the individual components. The relation is clear for the global galaxy and holds for both of the individual components, though it is considerably stronger for the bulges than the discs, which show a large amount of scatter. We propose that this can be explained by considering the bulge's position in the centre of the galactic potential well, making it easier to retain metals throughout its lifetime. In contrast, the disc will be more exposed to any external effects such as gas stripping through interactions and gas accretion.

\section{Kinematics} \label{kins}

In studying galaxy kinematics various parameters have been used to quantify the rotation of a galaxy, each with its own merit. In all cases the degree of rotation has been demonstrated to depend on various morphological properties, in particular with stellar mass and with bulge-to-total ratio, or proxies such as S\'{e}rsic index and velocity dispersion $\sigma_e$ \citep{Romanowsky2012,Cortese2016,Krajnovic2013,Graham2018}. This dependence demonstrates that bulges and discs play a key role in determining the evolutionary path of a galaxy, a role which is hard to determine without understanding the kinematic properties of the individual components themselves. Using our decomposed kinematics we can, for the first time, extract and quantify the kinematic properties of the individual components for a large sample of galaxies, and therefore understand how they contribute to the global kinematics of the galaxy.

\subsection{Specific angular momentum}
The angular momentum $j_\ast$, where the projected value is defined as 
\begin{equation} \label{eq:ang}
j_{\rm p}=\frac{\sum {F_{\rm i}  R_{\rm i}  \mid V_{\rm ilos} \mid}} {\sum F_{\rm i}},
\end{equation}
has been shown to depend on the stellar mass of the galaxy \citep{Fall1983}, with a scatter that depends on various morphological parameters, such as S\'{e}rsic index and bulge-to-total ratio \citep{Fall1983,Cortese2016}. \citet{Romanowsky2012} hypothesised that the location of a galaxy on the $M_\ast - j_\ast$ plane is therefore determined by the relative contributions of the bulge and disc, each of which should, in theory, follow their own distinct $M_\ast - j_\ast$ relation. They, and more recently \citet{Fall2018}, used strong simplifying assumptions to estimate $j_\ast$ values for bulge and disc components within their galaxies which indicated that the components do indeed follow their own $M_\ast - j_\ast$ relations, with bulges offset to lower $j_\ast$ values, similar to elliptical galaxies, and discs lying in similar regions to late-type disc galaxies. This was also explored by \citet{Rizzo2018}, who decomposed bulges and discs of 10 S0 galaxies in order to compare the $M_\ast - j_\ast$ relation of their discs to those of spiral galaxies, which they found to be very similar.  With our decomposed kinematics we can extend this analysis to determine more rigorously, for a large sample of galaxies, where bulges and discs lie on the $M_\ast - j_\ast$ plane. 

\begin{figure*}
    \centering
    \begin{tabular}{cc}
    \subfloat{\includegraphics[width=0.5\linewidth]{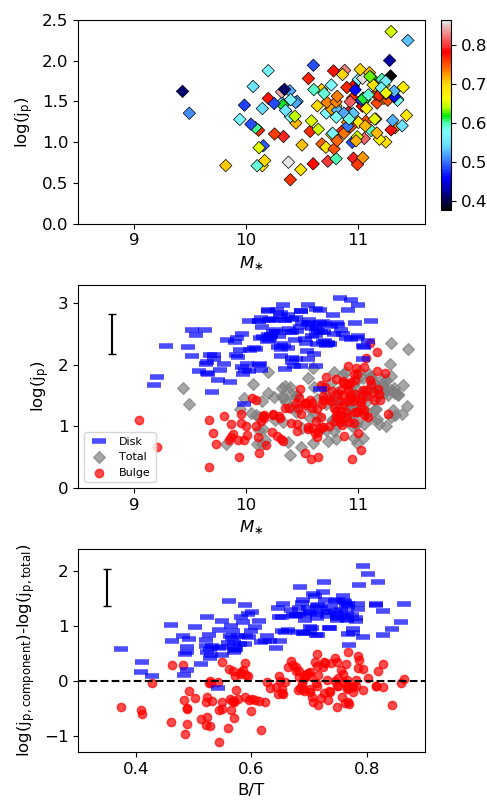}}&
    \subfloat{\includegraphics[width=0.445\linewidth]{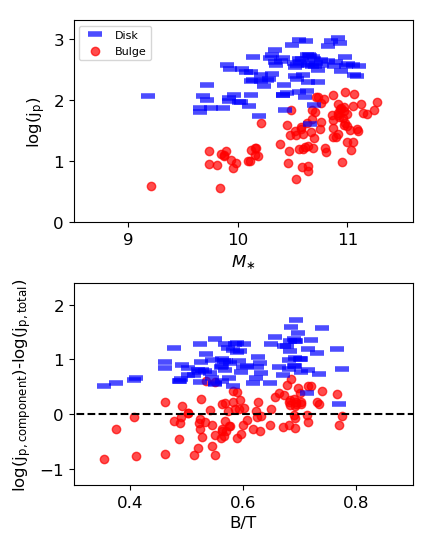}}\\
    \end{tabular}
    \caption{Top: angular momentum vs. stellar mass for global galaxy, coloured by bulge-to-total ratio. Middle:  angular momentum vs. stellar mass for decomposed bulge and disc and global galaxy, where the stellar mass of the individual components is approximated by $M_\ast \times B/T$ for the bulge and $M_\ast \times (1-B/T)$ for the disc.
    The left-hand panel shows the values for the bulge and disc calculated within 1$\rm R_e$.
    The relations for both bulge and disc are significant with p-value $ =(1.7\times10^{-13}, 3\times10^{-9})$ and r-value $=(0.57, 0.47)$ for bulge and disc respectively. 
    The right-hand panel shows the same relations but with the values calculated for overlapping bulge and disc regions, both for the individual component values and for the global values.
    Bottom left: the difference between the global values of $j$ and the component values for varying bulge-to-total ratio. Errors are estimated from the scatter in $j_{bulge}-j_{total}$ at large B/T, which for such bulge-dominated galaxies should be $\sim 0$.
    Bottom right: the same as bottom left but calculated across overlapping bulge and disc regions.}
    \label{fig:jmass}
\end{figure*}

Figure \ref{fig:jmass} demonstrates how the position of a galaxy  on the $M_\ast - j_\ast$ plane varies with bulge-to-total ratio, with lower B/T galaxies tending to have higher values of $j_\ast$ at a given mass. The dependence of the scatter on B/T is not as prominent as found for previous studies, most likely due to the relatively small range of B/T in our sample, but is still significant, with $j_\ast$ correlated with B/T with a p-value $=0.002$. It also shows the distinct $M_\ast - j_\ast$ relations when bulges and discs are decomposed, both when $j_\ast$ is determined within 1$\rm R_e$ as well as in overlapping regions. There is a clear offset between bulge and disc, with the global galaxy lying between the two components, though since these are early-type galaxies the majority have a high bulge fraction, resulting in the global values tending to lie much closer to those of the bulge than the disc. 

For the values of $j_\ast$ calculated in the overlapping regions, the offset between bulge and disc is slightly smaller, with average $\log(j_{Disk}) - \log(j_{Bulge})=0.87$, compared to $1.21$ when using all bins. This is due to the disk $j_\ast$ decreasing as the outer regions of high rotation tend not to be included, while the bulge values are not significantly altered since the excluded inner bins are not rotation dominated. While this implies that a small part of the separation in the 1$\rm R_e$ values is due to aperture effects, it is still clear that the components have systematically different angular momentum.
The discs lie within the same region as found by previous studies for late-type spiral galaxies \citep{Fall1983,Cortese2016}, i.e. galaxies with very low bulge-to-total ratios, as well as the separated discs in \citet{Romanowsky2012}. 

The relatively tight correlation in the discs of these galaxies, some of which are classed globally as slow rotating galaxies, demonstrates the remarkable consistency of the properties of these discs across all morphologies. There is a fairly large scatter in the relations for both components, which is expected as this is the projected angular momentum rather than the intrinsic angular momentum, resulting in a scatter purely due to the range in inclination. 
We do not attempt to correct for inclination here as this requires knowing the intrinsic axial ratio, q$_0$, of the components, a property which is highly uncertain.

The bottom panel of Figure \ref{fig:jmass} demonstrates clearly how the global angular momentum varies according to the contribution of the components. For low B/T the global value of $j_\ast$ is close to that of the disc, and as the B/T increases it moves closer to the value of the bulge. 
Due to difficulties inherent in a complex method such as this, there is some additional scatter in the values of $j_\ast$ for the individual components, which is why $j_\ast$ of the bulge in high B/T galaxies is occasionally higher than that of the global galaxy. This is not a physical phenomenon, and is just due to the difference between the bulge and global values of $j_\ast$ scattering around zero for the very bulge-dominated galaxies.

\begin{figure*}
    \centering
    \begin{tabular}{cc}
    \subfloat{\includegraphics[width=0.45\linewidth]{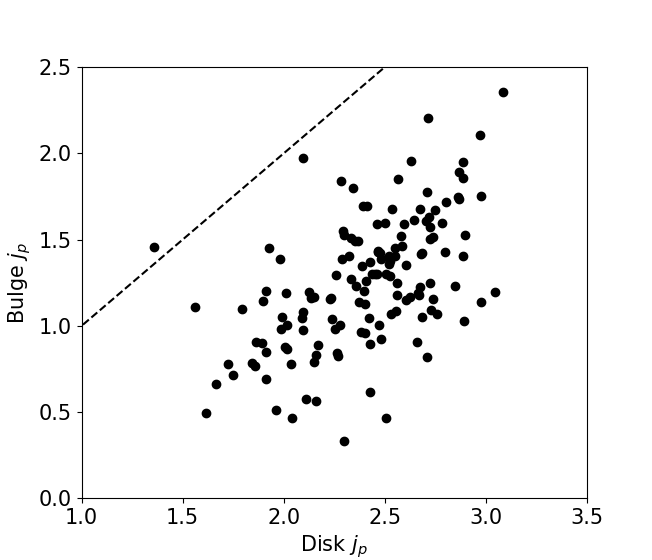}}&
    \subfloat{\includegraphics[width=0.44\linewidth]{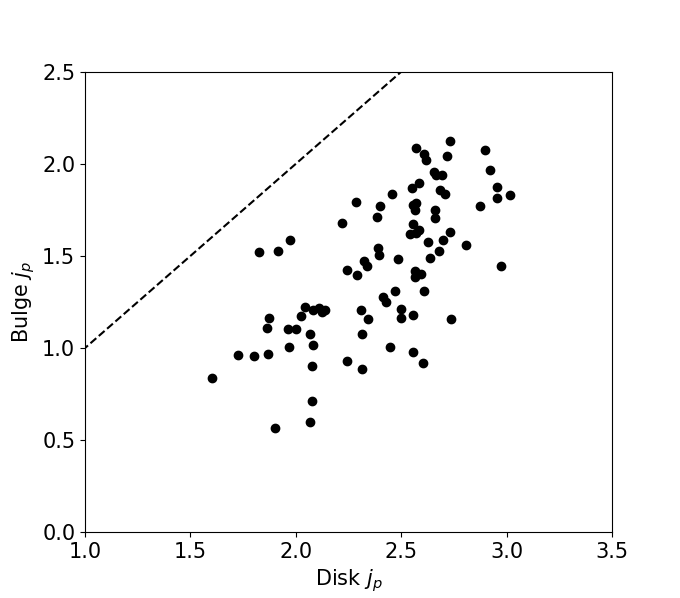}}\\
    \end{tabular}
    \caption{Projected specific angular momentum of the bulge and disc components, correlated with r-value = 0.55, p-value=2.3$\times10^{12}$ from a linear regression fit. The dashed line shows a one-to-one correlation. The left hand panel shows the relation for $j_\ast$ determined within 1$\rm R_e$, while the right hand panel shows the relation for $j_\ast$ within the overlapping regions.}
    \label{fig:jbd}
\end{figure*}

The angular momentum of the components is also clearly correlated, albeit with a large scatter, as shown in Figure \ref{fig:jbd}. This demonstrates a coupling of the components, suggesting that, as indicated by the stellar populations in Section \ref{pops}, the evolution of the two components is not independent. 

Interestingly, we find no clear correlation between the kinematics of bulges and their Sersic index, a parameter often used as a proxy for kinematics in determining how `discy' a bulge is. This agrees with previous work by \citet{Mendez2018}, who compared photometric and kinematic properties of field S0 galaxies in the CALIFA survey, and concluded that photometric parameters are not a good proxy for kinematics.

\subsection{Stellar Spin Parameter, $\lambda_R$}

\begin{figure*}
    \centering
    \includegraphics[width=\linewidth]{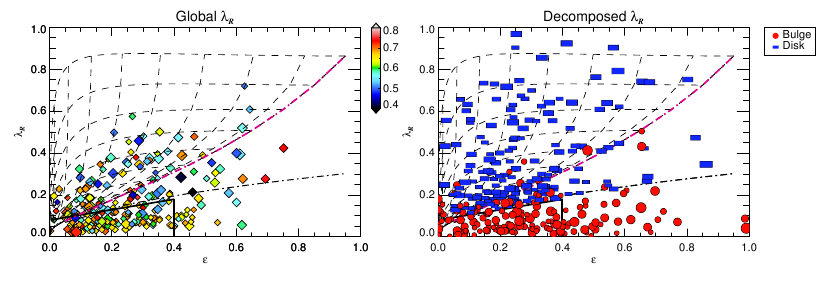}
    \caption{Stellar spin parameter $\lambda_{Re}$ vs ellipticity, $\epsilon$, for the global galaxy on the left, coloured by bulge-to-total ratio, and for the decomposed bulge and disc on the right. The blue bars correspond to disc components and the red circles correspond to bulges. The dashed contours correspond to dynamical models with $\delta=0.7 \times \epsilon$, as described in \citet{Cappellari2007}, for varying intrinsic ellipticities and inclinations, with the magenta line corresponding to edge-on galaxies. The dot-dashed line and solid black line correspond to the slow-fast rotator divide as defined by \citet{Emsellem2007} and \citet{Cappellari2016} respectively. The decomposed bulges tend to lie at low values of $\lambda_{Re}$, while discs lie at higher values. The global values of the galaxy are determined by the combination of the two components. Note that since the bulges and discs sit within the global potential of the galaxy, they can not be directly mapped to intrinsic dynamical properties in the same way as for galaxies as a whole. This figure is included to illustrate the relative positioning of the components, not to provide a detailed interpretation of their dynamical properties.}
    \label{fig:lam}
\end{figure*}

A related quantity, the stellar spin parameter, $\lambda_R$, defined in Equation \ref{eq:lam}, is a useful way of determining the relative contributions of rotational and random motions in a galaxy. The location of a galaxy on the $\lambda_R - \epsilon$ plane has been shown to depend on the bulge-to-total ratio of the galaxy \citep{Krajnovic2013}, demonstrating that, as for the $M_\ast-j_\ast$ relation, bulges and discs play a key role in determining the global properties of a galaxy. This effect is shown in the left panel of Figure \ref{fig:lam} for this sample. Using the decomposed kinematic maps we can also calculate this parameter for the individual components. Figure \ref{fig:lam} shows $\lambda_R$, within one $R_e$ ($\lambda_{Re}$), for each component, plotted against ellipticity, obtained for the individual components from the GALFIT decomposition. The contours correspond to the locations of galaxies based on dynamical models for different intrinsic ellipticities and inclinations. The values of $\lambda_{Re}$ have not been corrected for seeing, a correction which results in a relatively moderate increase in $\lambda_{Re}$. This increase is detailed in \citet{Graham2018}, though there is a large uncertainty in these values.

It is clear that when the components are separated the discs tend to occupy regions of higher $\lambda_R$, whereas the bulges tend to sit in lower regions, towards the slow-rotator region. Interestingly, many slow-rotator galaxies still show signs of a disc, the existence of which agrees with work by \citet{Krajnovic2013} who compared photometric decompositions with 
the kinematic signatures in early type galaxies from the ATLAS$^{\rm 3D}$  sample and found that 29\% of slow 
rotators still appear to contain an exponential component. They demonstrate that there appears to 
be a transitional region between fast and slow rotators, including slow rotators that contain an exponential disc \citep[e.g.][]{Li2018}, and fast rotators that are not described by an exponential profile. 

\begin{figure}
    \centering
    \includegraphics[width=\linewidth]{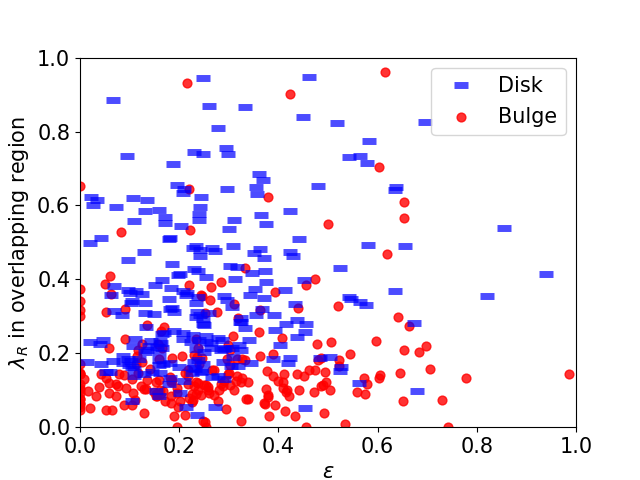}
    \caption{Same as for right hand panel of Figure \ref{fig:lam} but with stellar spin parameter $\lambda_{R}$ calculated within overlapping regions of bulge and disc to allow a direct comparison. While the bulges and discs are more widely distributed than in the Figure \ref{fig:lam}, discs still clearly tend to lie towards higher values of $\lambda_R$ than bulges.}
    \label{fig:lamov}
\end{figure}

The majority of discs sit within the contours showing the locations of dynamical models, satisfying the anisotropy condition, $\delta=0.7 \times \epsilon$ \citep{Cappellari2007} for varying intrinsic ellipticities and inclinations. While the global kinematics do not neatly fit into a particular dynamical model, being distributed across the slow/fast-rotator divide, the individual components for the most part do. Some caution is necessary in the interpretation of this figure; since bulges and discs are not self-gravitating systems, their location on the $\lambda_R-\epsilon$ plane cannot be mapped directly to their intrinsic dynamical properties. This issue is addressed in the following section \ref{jammod}, and here we simply use the figure as a way to illustrate the separation of the components. 
Figure \ref{fig:lamov} shows the $\lambda_R-\epsilon$ distribution for values calculated over the overlapping bulge and disc regions. 
The values for both components are more widely distributed than in Figure \ref{fig:lam}, which is expected as they include less data, but the discs still clearly lie at higher values of $\lambda_R$ than the bulges, with an average difference of $\lambda_R \sim 0.2$, showing that the difference in the components is not simply due to the difference in radial range over which they are determined.
Dynamical modelling provides an additional way of accounting for these problems, allowing extraction the intrinsic kinematic properties of the components, irrespective of radius. The results of this are presented in Section \ref{jammod}.

Similar to the location of a galaxy $M_\ast - j_\ast$ relation, this difference in the properties of the components again points to the global properties of a galaxy being determined by the relative contribution of the bulge and disc. Separating out these components may therefore provide a more complete way of dynamically modelling galaxies.

\subsection{Dynamical modelling} \label{jammod}

Treating bulges and discs as distinct and separable components is actually not entirely correct due to the fact that they are not self-gravitating systems, but rather sit within the global potential of the galaxy. Mapping the location of a component on the $\lambda_{Re}-\epsilon$ plane, as shown in Figure \ref{fig:lam}, to its dynamical properties is therefore not as simple for individual components as for the galaxy as a whole. In addition, due to the calculated $\lambda_{Re}$ only including bins where the component fraction is greater than 30\%, the central regions of the disc and the outer regions of the bulge do not contribute. This makes direct comparisons between the components difficult. Therefore, to try more rigorously to understand the dynamical properties of these components, we have used a simple dynamical model to describe the individual component kinematics modelled within the global potential of the galaxy. In addition, by comparing the kinematic components acquired through dynamical modelling to those determined by spectroscopic decomposition, we can check that these two decomposition methods are consistent.

To produce dynamical models of the individual components, we use the Jeans Anisotropic Modelling method (JAM), an anisotropic generalization of the semi-isotropic (two-integral)
axisymmetric Jeans formalism of \citet{Cappellari2008}, and apply it to a subset of the sample of spectroscopically decomposed galaxies, selecting 50 galaxies with the highest signal-to-noise and cleanest rotation fields, chosen by eye, for which such complex modelling is suitable.

To model the components we require the distribution of light and mass of each component within the galaxy. These are derived using the Multi-Gaussian Expansion (MGE) method \citep{Emsellem1994} with software developed by \citet{Cappellari2002} to fit the bulge and disc images created by GALFIT during a photometric decomposition, as described in Section \ref{phot}. This surface brightness is then deprojected to produce the luminosity density, and, assuming a constant mass-to-light ratio, this is then converted to a stellar mass distribution. Since both components sit in the global potential of the galaxy, the mass distributions of both components were combined to form a global potential.

The JAM fitting consists of two steps; first, with the luminosity density obtained, the galaxy $V_{\rm rms} = \sqrt{V_{\rm los}^{2}+\sigma^{2}}$ can be fully described by the anisotropy parameter $\beta_{\rm z}=1-(\sigma_{\rm z}/\sigma_{\rm R})^2$, the inclination and the mass-to-light ratio. We find the best fitting combination of these parameters by minimizing the residuals $(V_{\rm  rms})_{\rm obs} - (V_{\rm rms})_{\rm model}$ using a least-squares fit.

Second, using these parameters a model for the line-of-sight velocity, $V_{\rm los}$, can be constructed, described by a parameter, $\kappa$, the amount by which an isotropic velocity field of $\sigma_{\rm \phi}=\sigma_{\rm R}$ must be scaled by to fit the data. A large value of $\kappa$ implies faster rotation and $\kappa=0$ implies no rotation. Assigning different values of $\beta_{\rm z}$ and $\kappa$ to the bulge and disc luminosity densities allows us to parameterise the components individually. However, in general measuring the velocity dispersion of a spectrum is more difficult than measuring $V_{\rm los}$. This introduces additional uncertainties in the modelling of $V_{\rm rms}$ and allowing for two different values of $\beta_{\rm z}$ only increases the degeneracies of the fit. We therefore only allow for two components in the fit to $V_{\rm los}$, i.e. allowing for two values of $\kappa$. Since the value of $\beta_{\rm z}$ used in the fitting of $V_{\rm los}$ does not have a significant effect on the value of $\kappa$ obtained, we use the single global value of $\beta_{\rm z}$ for both components in the fits to $V_{\rm los}$.

To determine the dynamical properties of the bulge and disc, we construct a model using two different $\kappa$ values for the luminosity density of the bulge and disc. As for the single component model, we find the best fitting model of $V_{\rm los}$ for a given $\kappa_{\rm bulge}$ and $\kappa_{\rm disc}$ by minimising the residuals $(V_{\rm los})_{\rm obs} -(V_{\rm los})_{\rm model}$ using a least-squares fit.

All the galaxies in the sample were visually inspected to ensure that the models, both for a single component and two components, were a good description of the data. Due to unsatisfactory photometric decompositions and complicated kinematics or the JAM fitting failing, a number of galaxies were discarded from the final sample, leaving a core of 29 dynamically-simple fast-rotating early-type galaxies.

One issue in the modelling process is that we do not take into account dark matter. In these central regions the effect should not be large, but potentially it could affect the outer disc--dominated region more than the inner bulge--dominated region, resulting in larger $\kappa$ values for the disc than the bulge. Our sample includes galaxies from both the primary and the secondary MaNGA sample, covering out to $1.5R_{\rm e}$ and $2.5R_{\rm e}$ respectively. If dark matter were having an effect, we would expect there to be an offset in $\kappa$ between the primary and secondary sample. However, we see no difference between these two samples and therefore conclude that dark matter is likely not having a significant effect on the results.

\begin{figure*}
    \centering
    \begin{tabular}{c:c}
    \textbf{\large{JAM decomposition}} & \textbf{\large{Spectroscopic decomposition}} \\
    \subfloat{\includegraphics[width=0.5\textwidth]{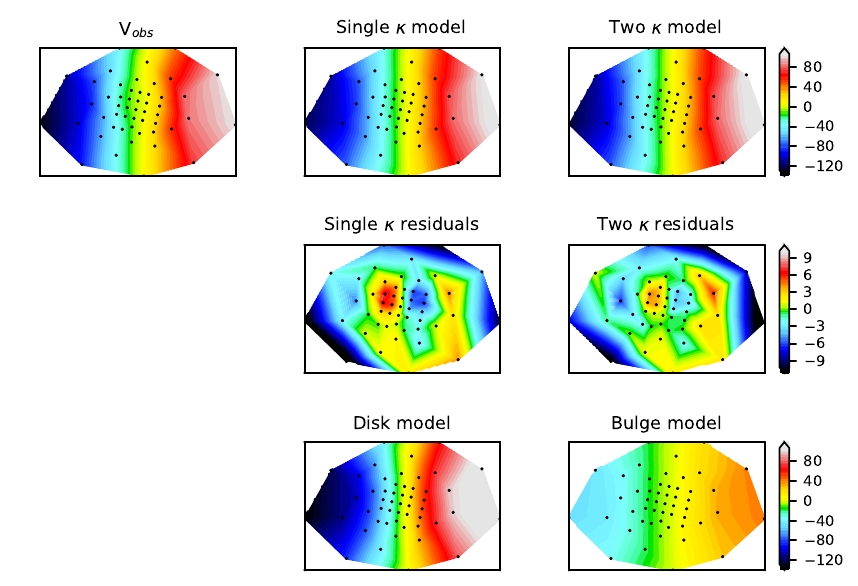}}&
    \subfloat{\includegraphics[width=0.4\textwidth]{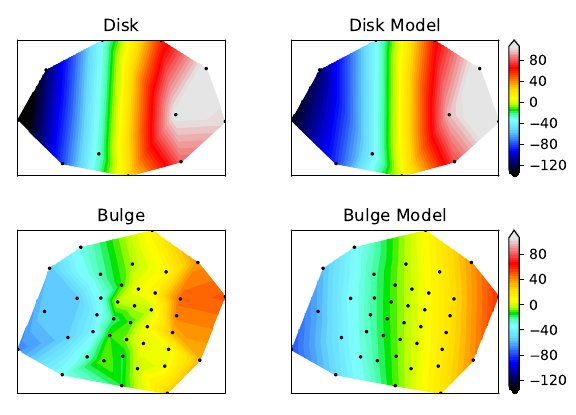}}\\
    \subfloat{\includegraphics[width=0.5\textwidth]{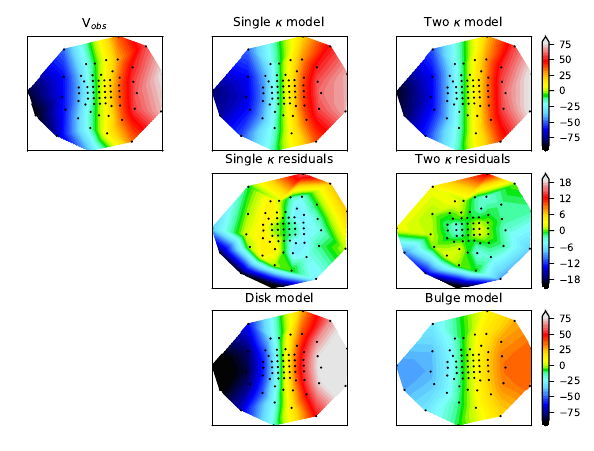}}&
    \subfloat{\includegraphics[width=0.4\textwidth]{8139-6102-jam-bd.png}}\\
    \end{tabular}
        \caption{On the left are the kinematic maps and the best fitting JAM models for two galaxies in the sample, fitting with a single component and with two components. On the right are the maps from the spectroscopic decomposition and their best fitting JAM models for the same two galaxies. The black points show the locations of the bins. Note that the maps for the individual components only fit the points where the component contributes greater than 50\% of the flux, therefore they cover a smaller area than the maps of the global velocity field. The velocity maps for the components obtained through dynamical modelling agree well with those obtained through spectroscopic decomposition. \label{fig:jams}}
 \end{figure*}

To compare the kinematics of the components found though the JAM modelling to those obtained through spectroscopic decomposition, we repeat the above process, but fitting the individual JAM model component velocity fields to those obtained from the spectroscopic decomposition. This gives a second value of $\kappa$ for both the bulge and the disc, which we can compare to the values from the fits to the global galaxy velocity field. To ensure that the kinematics are correct we only use bins where the flux of the component is greater than 50\% of the total flux. The velocity maps and resulting JAM models for two galaxies in the sample are shown in Figure \ref{fig:jams}, demonstrating that the velocity maps of the components from the spectroscopic decomposition, shown in the right-hand panel, agree well with those from the dynamical modeling shown in the left-hand panel.

\subsection{Comparing spectroscopic decomposition to dynamical modelling}

\begin{figure}
    \centering
    \includegraphics[width=0.5\textwidth]{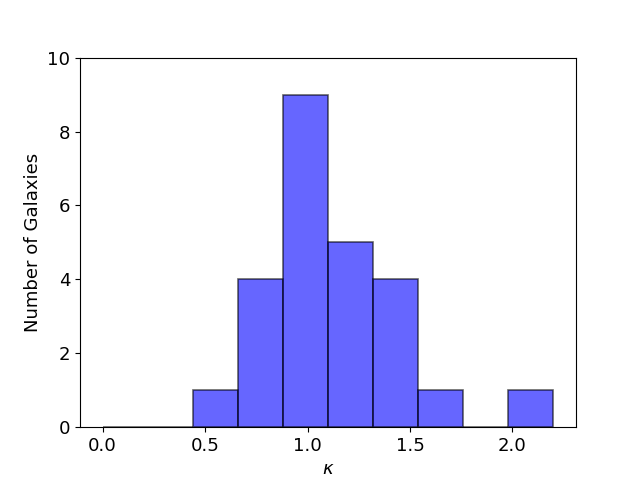}
    \caption{Histogram of $\kappa$, the ratio of the observed velocity to the velocity field predicted by a JAM model with  $\sigma_{\phi}=\sigma_{R}$, for a single component JAM model. The distribution of these values is consistent with \citet{Cappellari2016}, finding that fast-rotator galaxies tend to have $\kappa \sim 1$. \label{fig:kappahist}}
\end{figure}

\begin{figure*}
    \centering
	\includegraphics[width=0.8\linewidth]{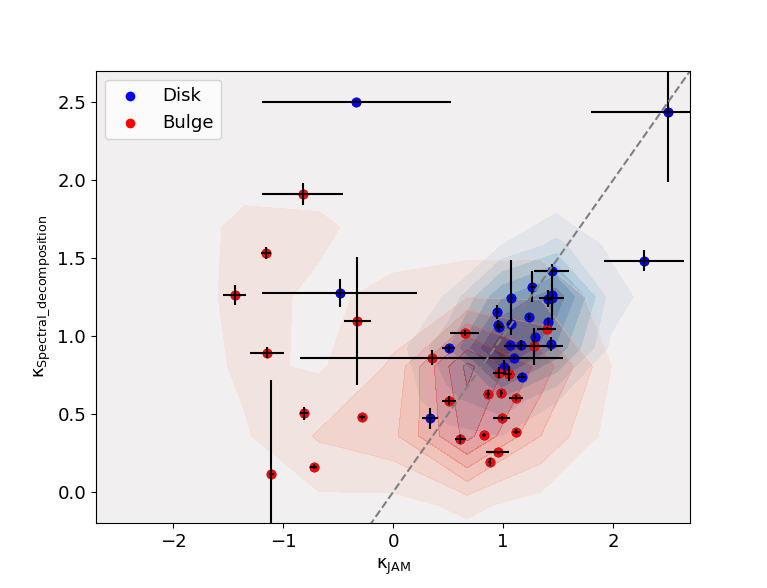} 
    \caption{Comparison between $\kappa$ values obtained through JAM modelling of the global velocity field, and those obtained through JAM modelling of the spectroscopically decomposed components. Error-bars correspond to $1-\sigma$ errors calculated by the minimization function. Contours show the density of points. The $\kappa$ values obtained through the two decomposition methods are consistent, and the agreement in finding an offset in $\kappa$ values for the bulge and disc demonstrates that bulges and discs have systematically different dynamical properties. However, these differences are not large and both the bulges and discs have $\kappa$ values consistent with those of fast-rotator galaxies considered as a whole. \label{fig:kappacom}}
\end{figure*}

The values of $\kappa$ using a single component model are shown in Figure \ref{fig:kappahist}. The peak at $\kappa=1$ agrees with \citet{Cappellari2016}, who found that a model with an oblate velocity ellipsoid ($\sigma_{\rm \theta} = \sigma_{\rm R}$) is a good description of the kinematics of most fast rotators.

Figure \ref{fig:kappacom} shows the $\kappa$ parameter for the bulges and discs, both from the fits to the global velocity field, and from the fits to the spectroscopically decomposed components. For the majority of galaxies the values of $\kappa$ show good agreement between the two decomposition methods. Notably, while $\kappa_{\rm disc}$ tends to be around, or slightly greater than, 1, $\kappa_{\rm bulge}$ tends to be less than 1. It is interesting to compare these values to those found for regular and non-regular rotators in \citet{Cappellari2016}, with regular rotators tending to have $\kappa \sim 1$ while non-regular rotators have $\kappa \sim 0$. Bulges, while having lower values of $\kappa$ than the discs, still have higher $\kappa$s than slow-rotators, and for the most part are consistent with the lower tail end of the distribution of $\kappa$ for fast-rotator galaxies. 
It is important to note that the JAM modelling has been performed on a sub-sample of the galaxies which have been spectroscopically decomposed, and we can not therefore conclude that is supports the entire technique of spectroscopic decomposition. However, the agreement thus far is encouraging and hopefully in the future it will be possible to extend the comparison to a larger number of galaxies.

There is also a group of galaxies that show a negative $\kappa$ for the bulge component in the JAM fits to the global velocity field, implying counter--rotation. There is, however, no indication of a counter rotating component from the observed kinematic maps or from the spectroscopically decomposed components, making it likely that this is not a physical phenomenon, but an effect of the JAM modelling process failing to account for models that deviate significantly from the simple case of an oblate velocity ellipsoid at all radii. 
JAM modelling is also not suitable for modelling slow-rotating galaxies. Since many of the bulges show significantly less rotation than the discs, the counter-rotation may be a result of the models failing to account for these less rotationally-supported systems.
While this simple modelling approach is sufficient for a crude comparison, a more sophisticated dynamical model, for example Schwarzchild modelling, or higher resolution data, would be required to really determine the underlying physics of these systems.

\subsection{Addressing the distinctness of JAM components}

Photometric decompositions and, more recently, spectroscopic decompositions have demonstrated the existence of multiple components within galaxies. However, previous work has found that fast-rotator early-type galaxies are well represented by a single dynamical model, suggesting that bulges and discs have similar anisotropies. An interesting question to therefore address is whether the two distinct kinematic components have different intrinsic kinematic properties, i.e. whether a two-component JAM model actually improves the fit to the data, or whether it introduces unnecessary complexity to a system sufficiently described using a single component model.

Having distinct kinematics does not necessarily mean the anisotropy of the components is different; for example the ratio of $\sigma_{z}/\sigma_{R}$ could be the same for bulges and discs, leading to the same value of $\beta_{z}$, even if the values of $\sigma_{z}$ and $\sigma_{R}$ are very different, as is expected given the well accepted fact that bulges have higher velocity dispersion than discs. However, given the different formation scenarios for bulges and discs, and the myriad of processes which can effect the orbits of the stars within them, it would somewhat surprising, and very interesting, if this ratio is indeed the same for both components.

As a way of testing whether the two components are dynamically distinct we compared the single-component and two-component models to the data. Since including additional degrees of freedom in a fit will inevitably lead to an improvement in chi-squared, determining when the inclusion of a second component corresponds to an actual physical component is not easy. As a criterion for the two-component model showing significant improvement compared to the single-component model, we determined whether the two-component model was closer to the data than it was to the single-component model, i.e. whether:
\begin{equation}
\sum^{nbins} (V_{\rm 2comp\_model} - V_{\rm obs})^{2} < \sum^{nbins} (V_{\rm 2comp\_model}- V_{\rm 1comp\_model})^{2}.
\end{equation}
If this is the case the improvement is most likely due to the two-component model more accurately describing the kinematics of the galaxy rather than just the additional degrees of freedom allowing for small adjustments in the model. 

Interestingly, none of the galaxies in our sample fulfill this criterion, therefore implying that bulges and discs do not have different anisotropies. However, due to the un-physical kinematics in several of the dynamical models, indicating that this simplistic model may not be sufficient in describing more complex systems, we feel that there is insufficient evidence at this point to conclude that bulges and discs are dynamically indistinct. In addition, while including two components does not lead to a statistically significant improvement in the model, the systematic difference in bulge and disc $\kappa$ is a strong indication of an underlying difference in their properties. This test illustrates the difficulties and degeneracies in building dynamical models for multiple components. Higher resolution data and more detailed dynamical models such as those used by \citet{Zhu2018} may be necessary to truly understand the kinematics of the individual components.

\section{Discussion} \label{diss}

\subsection{Applying spectroscopic decomposition to large IFU surveys}

In this paper we have demonstrated that spectroscopic decomposition can be successfully applied to large IFU surveys. However, applying any complex method, such as the one used here, to large numbers of galaxies is not a simple task. As demonstrated by the significant number of galaxies cut from the original sample, the decomposition can not be applied to all galaxies, and will not always be successful. For the decomposition to work it requires high signal-to-noise data, and sufficient flux in both bulge and disc components.

\begin{figure}
    \centering
    \includegraphics[width=0.5\textwidth]{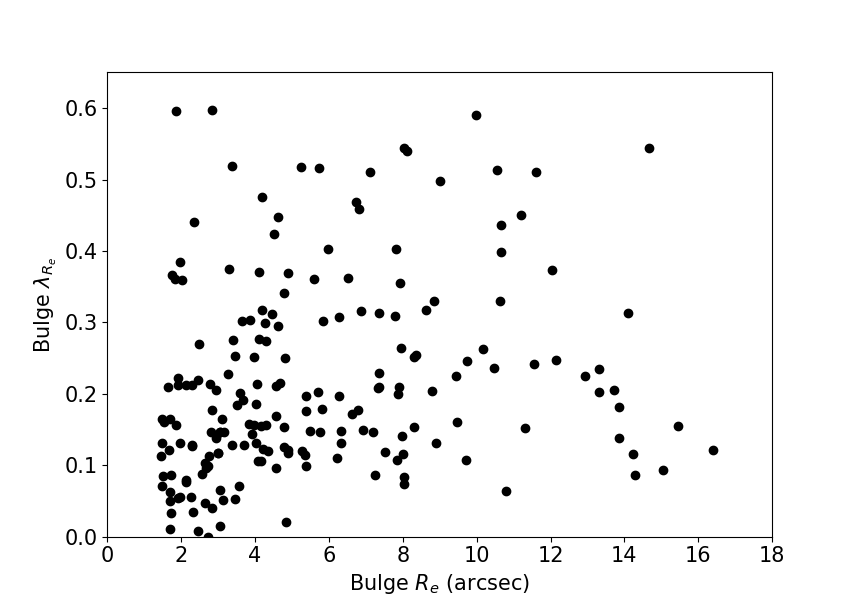}
    \caption{Bulge effective radius versus bulge $\lambda_{Re}$, showing that the kinematics can be extracted even for bulges with effective radii comparable to the FWHM of MaNGA data. \label{fig:reff}}
\end{figure}

A particular concern when applying this method on large IFU surveys is how the spatial resolution can affect the properties of the bulge. The cuts made on the sample remove components which have very few bins, and therefore remove the smallest bulges, however, there are still bulges in the sample which have effective radii comparable to the FWHM of MaNGA observations. If this is having a significant effect on the resulting kinematics then we would expect that all bulges below a certain effective radius would have systematically lower angular momentum than larger bulges, due to the kinematics being smeared out. Figure \ref{fig:reff} shows how the $\lambda_{Re}$ values of the bulge vary with effective radius. While the lowest angular momentum bulges with $\lambda_{Re} < 0.1$ do tend to be fairly small, there are also many equally small bulges with high values of $\lambda_{Re}$, demonstrating that even in bulges that are not fully resolved the kinematics can still be extracted. 
While we cannot completely rule out the effects of spatial resolution, it is clear that bulges of all sizes show consistently lower angular momentum than their corresponding discs, demonstrating that the kinematic differences between the components are not determined by spatial resolution effects.

While some care must be taken when applying this method to large samples of galaxies, providing the quality of data, the signal-to-noise and the morphology of the galaxies are suitable, and the resulting parameters are carefully checked, spectroscopic decomposition can reliably be applied to large IFS surveys.

\subsection{Stellar populations and kinematics of bulges and discs}

Separating out galaxies into central and extended components has long been used as a way to explore their formation histories. In \citet{Tabor2017} we considered whether these components represented two different stellar populations with distinct kinematics, which we found to be the case. Now with a larger sample of early-type, barless galaxies, we extend this analysis from whether they are different to \textit{how} they are different. 

The stellar populations of the components show that bulges and discs of these high mass galaxies in general have similar ages, but the bulges have systematically higher metallicities. For both age and metallicity, the discs show a wider range of values, reaching significantly lower metallicities and ages than the bulges. This is most likely because discs tend to have more varied star-formation histories, being more susceptible to external processes such as gas accretion, suppression or removal \citep{Smethurst2015}. The position of the bulge in the centre of the galactic potential well also allows metal-enriched gas to be retained, and higher metallicity stellar populations to be formed. This also explains the stronger mass-metallicity relation in the bulge compared to the disc.

Where galaxies lie on the $\lambda_{\rm R} - \epsilon$ diagram is dictated by the intrinsic dynamical properties of the galaxy. However, the global values of $\lambda_R$ tend not to correspond neatly to any particular simple dynamical model. This is unsurprising when you consider that these galaxies are made up of multiple components, the effect of which is clear when looking at how the position of a galaxy depends on its bulge-to-total ratio. To truly understand the kinematics of these galaxies we must therefore separate them into their constituent components. When we do this, we find that bulges and discs have very different $\lambda_R$ parameters, with discs tending to be distributed towards higher angular momentum than bulges, which lie in, or close to, the slow rotator region. 
By combining the spectroscopic decomposition with dynamical modelling we find that while discs tend to have $\kappa \gtrsim 1$, bulges appear to have $\kappa \lesssim 1$, implying that there are systematic differences between bulges and discs. 

This parameterisation of the kinematics of the components could prove very useful when compared across galaxy morphologies. For instance, similarities and differences between the bulges of star-forming versus quenched galaxies could be key in understanding whether the quenching occurred via internal secular processes, or external processes such as mergers [see \citet{Kormendy2004} for a review].

The angular momentum $j_\ast$ of the components can also give insight into their history. When decomposed, each component follows a clear $M_\ast - j_\ast$ relation, which for the discs is the same relation as found for late-type disc galaxies, while the bulges sit on, or just below, the global values for slow-rotator elliptical galaxies. This demonstrates a remarkable consistency in these components across all morphologies, and is a persuasive indicator that not all slow-rotator galaxies can be classified as being fundamentally different from the more clearly discy lenticular galaxies, agreeing with observations previously made in regards to photometry \citep{Krajnovic2013} and star-formation histories \citep{Smethurst2018}. Combined with the coupling of the kinematics between components, this consistency also indicates that whatever the mechanism that leads to the transformation of these galaxies, it is one which evidently either retains the underlying kinematics of the individual components, or rebuilds the components in a very similar manner. Progress is being made using simulations to understand the effects that different evolution paths have on the final structure of an early-type galaxy, both through secular processes and through mergers \citep{Barnes2002,Springel2005,Robertson2006,Martig2009,Hopkins2009,Moster2011,Querejeta2015,Querejeta2015a,Saha2018}. However, until simulations have demonstrated that the consistent and separable kinematic relations of both bulges and discs can be replicated in merger remnants, it will be unclear as to whether they provide a realistic method of producing the early-type galaxies observed here.

The great advantage of this method of spectroscopic decomposition is in the extraction of both the kinematics and the stellar populations, which when combined provide a detailed view into the evolution of a galaxy. What is intriguing about the outcome of this synthesis is that there appears to be little clear correlation between the stellar composition of the components and their kinematics. In particular, where a component lies on the $M_\ast - j_\ast$ plane does not depend on the age of that component, or the relative ages of the components. While there is a dependence of metallicity on angular momentum, it is much stronger for, and therefore likely determined by, the stellar mass. While the different components are therefore clearly dynamically distinct, their stellar populations are often quite similar. 
However, the range in ages in the sample is small, being all early-type galaxies and therefore not including young discs, and discriminating between populations which are uniformly old is not easy.
Extending this work to include star-forming galaxies may be the key in truly understanding exactly where and when star formation is switched off, and the effect, if any, this process has on the kinematics of the components.

\section{Conclusions} \label{conc}

In this paper we have spectroscopically decomposed a sample of MaNGA early-type galaxies into bulge and disc components. This allows us to obtain stellar populations and kinematic parameters for the components separately, and look at their similarities and differences. Our main conclusions are as follows:

\begin{itemize}
    \item Bulges and discs have similar ages, but bulges are systematically offset to higher metallicities. Discs tend to span a wider range of both age and metallicity, reaching much younger ages and lower metallicities, indicating a more varied star-formation history compared to the bulges.
\end{itemize}
The kinematic properties of galaxies, specifically the stellar spin parameter, $\lambda_R$, and the specific angular momentum, $j_\ast$, have been shown to depend on the bulge-to-total ratio of the galaxy. Decomposing the bulges and discs allow us to understand the root of this dependence, showing that:
\begin{itemize}
    \item  Bulges and discs follow their own $M_\ast-j_\ast$ relations, with bulges offset to lower values of $j_\ast$ compared to the discs. Where the galaxy as a whole sits on this relation is then determined by how much bulge and disc are present in the galaxy.
    \item A similar trend is seen for the position of bulges and discs on the $\lambda_R - \epsilon$ diagram, with bulges lying at lower values of $\lambda_R$ than the global galaxy, while discs move to higher values. 
\end{itemize}
By constructing dynamical models of a sub-sample of galaxies and allowing for different dynamical properties for the bulge and disc components we find:
\begin{itemize}
    \item The best-fit bulge and disc kinematic parameters from the dynamical models are consistent with those obtained through spectroscopic decomposition.
    \item The dynamics of bulges and discs are best described by slightly different values of $\kappa$, corresponding to different anisotropies, with discs showing similar, or slightly higher, values to those found for fast-rotator galaxies, while bulges have slightly lower values.
\end{itemize}

It is likely that no one evolutionary path applies to all early-type galaxies, even across similar morphologies and mass, but combining the stellar populations and kinematics of the components offers some critical insights into how each may be affected. Applying spectroscopic decomposition to a wide range of morphologies, mass and environment may provide the details necessary to untangle the effects of each.

\section*{Acknowledgements}

The authors would like to gratefully acknowledge Michele Cappellari for his advice on this work, and Eric Emsellem for useful discussions on the subject. We also thank the anonymous referee for helping to improve the paper. Funding for the  Sloan Digital Sky Survey IV has  been provided by the
Alfred P.  Sloan Foundation, the  U.S. Department of Energy  Office of
Science,  and the  Participating Institutions.  SDSS- IV  acknowledges
support and  resources from the Center  for High-Performance Computing
at the University of Utah. The SDSS web site is www.sdss.org.

SDSS-IV is  managed by the  Astrophysical Research Consortium  for the
Participating  Institutions of  the SDSS  Collaboration including  the
Brazilian Participation  Group, the Carnegie Institution  for Science,
Carnegie  Mellon  University,  the Chilean  Participation  Group,  the
French   Participation    Group,   Harvard-Smithsonian    Center   for
Astrophysics,  Instituto de  Astrof\'{i}sica  de  Canarias, The  Johns
Hopkins University, Kavli Institute for the Physics and Mathematics of
the Universe (IPMU) / University  of Tokyo, Lawrence Berkeley National
Laboratory,  Leibniz  Institut   f\"{u}r  Astrophysik  Potsdam  (AIP),
Max-Planck-Institut    f\"{u}r     Astronomie    (MPIA    Heidelberg),
Max-Planck-Institut     f\"{u}r     Astrophysik    (MPA     Garching),
Max-Planck-Institut f\"{u}r Extraterrestrische  Physik (MPE), National
Astronomical Observatory  of China,  New Mexico State  University, New
York University, University of Notre Dame, Observat\'{o}rio Nacional /
MCTI,  The  Ohio  State  University,  Pennsylvania  State  University,
Shanghai Astronomical Observatory, United Kingdom Participation Group,
Universidad  Nacional   Aut\'{o}noma  de  M\'{e}xico,   University  of
Arizona,  University  of  Colorado   Boulder,  University  of  Oxford,
University of Portsmouth, University  of Utah, University of Virginia,
University   of  Washington,   University  of   Wisconsin,  Vanderbilt
University, and Yale University.




\bibliographystyle{mnras}
\bibliography{paper3} 




\bsp	
\label{lastpage}
\end{document}